\documentclass[prl,twocolumn,superscriptaddress,floatfix]{revtex4}
\usepackage{graphics,epsfig,amsfonts,amssymb,amsmath,ulem,color}
\definecolor{laura}{RGB}{10, 150, 10}
\newcommand{\beginsupplement}{%
        \setcounter{table}{0}
        \renewcommand{\thetable}{S\arabic{table}}%
        \setcounter{figure}{0}
        \renewcommand{\thefigure}{S\arabic{figure}}%
     }

\begin{document}
\title{Strong correlations and the search for high-Tc superconductivity in chromium pnictides and chalcogenides}
\author{J.M. Pizarro}
\affiliation{Instituto de Ciencia de Materiales de Madrid, 
ICMM-CSIC, Cantoblanco, E-28049 Madrid (Spain).}
\author{M.J. Calder\'on}
\email{calderon@icmm.csic.es}
\affiliation{Instituto de Ciencia de Materiales de Madrid, ICMM-CSIC, Cantoblanco, E-28049 Madrid (Spain).}
\author{J. Liu}
\affiliation{Instituto de Ciencia de Materiales de Madrid, 
ICMM-CSIC, Cantoblanco, E-28049 Madrid (Spain).}
\affiliation{School of Physics, Shandong University, Jinan 250100, People's Republic of China}
\author{M.C. Mu\~noz}
\affiliation{Instituto de Ciencia de Materiales de Madrid, ICMM-CSIC, Cantoblanco, E-28049 Madrid (Spain).}
\author{E. Bascones}
\email{ leni.bascones@icmm.csic.es}
\affiliation{Instituto de Ciencia de Materiales de Madrid, ICMM-CSIC, Cantoblanco, E-28049 Madrid (Spain).}
\date{\today}
\begin{abstract}  
Undoped iron superconductors accommodate $n=6$ electrons in five d-orbitals. Experimental and theoretical evidence shows that the strength of correlations increases with hole-doping, as the electronic filling approaches half-filling with $n=5$ electrons. This evidence delineates a scenario in which the parent compound of iron superconductors is the half-filled system, in analogy to cuprate superconductors. In cuprates the superconductivity can be induced upon electron or hole doping. In this work we propose to search for high-Tc superconductivity and strong correlations in chromium pnictides and chalcogenides with $n<5$ electrons. By means of ab-initio, slave spin and multi-orbital RPA calculations we analyse the strength of the correlations and the superconducting and magnetic instabilities in these systems with main focus on LaCrAsO. We find that electron-doped LaCrAsO is a strongly correlated system with competing magnetic interactions, being $(\pi,\pi)$ antiferromagnetism and nodal d-wave pairing  the most plausible magnetic and superconducting instabilities, respectively.  
\end{abstract}
\maketitle

Since high-Tc superconductivity was discovered in iron based compounds, the search for superconductivity has been extended to materials with a similar lattice structure but a different d-element. This search has led to the discovery of a few new superconductors based on Ni, Pt, Ir, Rh or Pd, but their critical temperatures do not rise beyond a few Kelvin~\cite{ronning_jpcm2008,bauer_prb2008,ronning_prb2009,tomioka_prb2009,
kudo_jpsj2010,berry_prb2009,hirai_physicac2010,imai_ssct2013}.  On the other hand, Mn-based pnictides  are antiferromagnetic insulators~\cite{singh_prb09,emery_prb2011,satya_prb2011,simonson_pnas2012,beleanu_prb2013,lamsal_prb2013,mcguire_prb2016} when undoped and suppression of magnetism with pressure in LaMnPO has not resulted in superconductivity~\cite{guo_scirep2013}. Isostructural Cr-pnictides, significantly less studied, are antiferromagnetic metals~\cite{singh-prb2009,park-inchem2013,paramanik-prb2014,jiang-prb2015}.

A large part of the community believes that the proximity to an antiferromagnetic phase is a key ingredient to find unconventional high-Tc superconductivity. The role of the Fermi Surface and the electronic correlations is currently debated. The latter ones have been emphasized on cuprates, which are Mott insulators when undoped. However, the relevance of electronic correlations was questioned in iron superconductors due to the metallic character of their antiferromagnetic state. Nevertheless, significant mass enhancements have been measured in many iron compounds~\cite{nosotras_review2016,biermann-review2016,liu08-zhou, qazilbash09,zpyin-natmaterials2011,sebastian_review, mingyi-prl2013,terashima2013,nakajima-jpsj2014,mingyi-natcomm2015,mingyi-prb2015}. 
The essential difference between cuprates and iron based materials is the multi-orbital character of the latter. Cuprates are usually described with a single-orbital which is half-filled in the parent compound while the so-called undoped iron superconductors accommodate $n=6$ electrons in the five Fe d-orbitals, with an average filling per orbital of $1.2$. 

\begin{figure}
\leavevmode
\includegraphics[clip,width=0.5\textwidth]{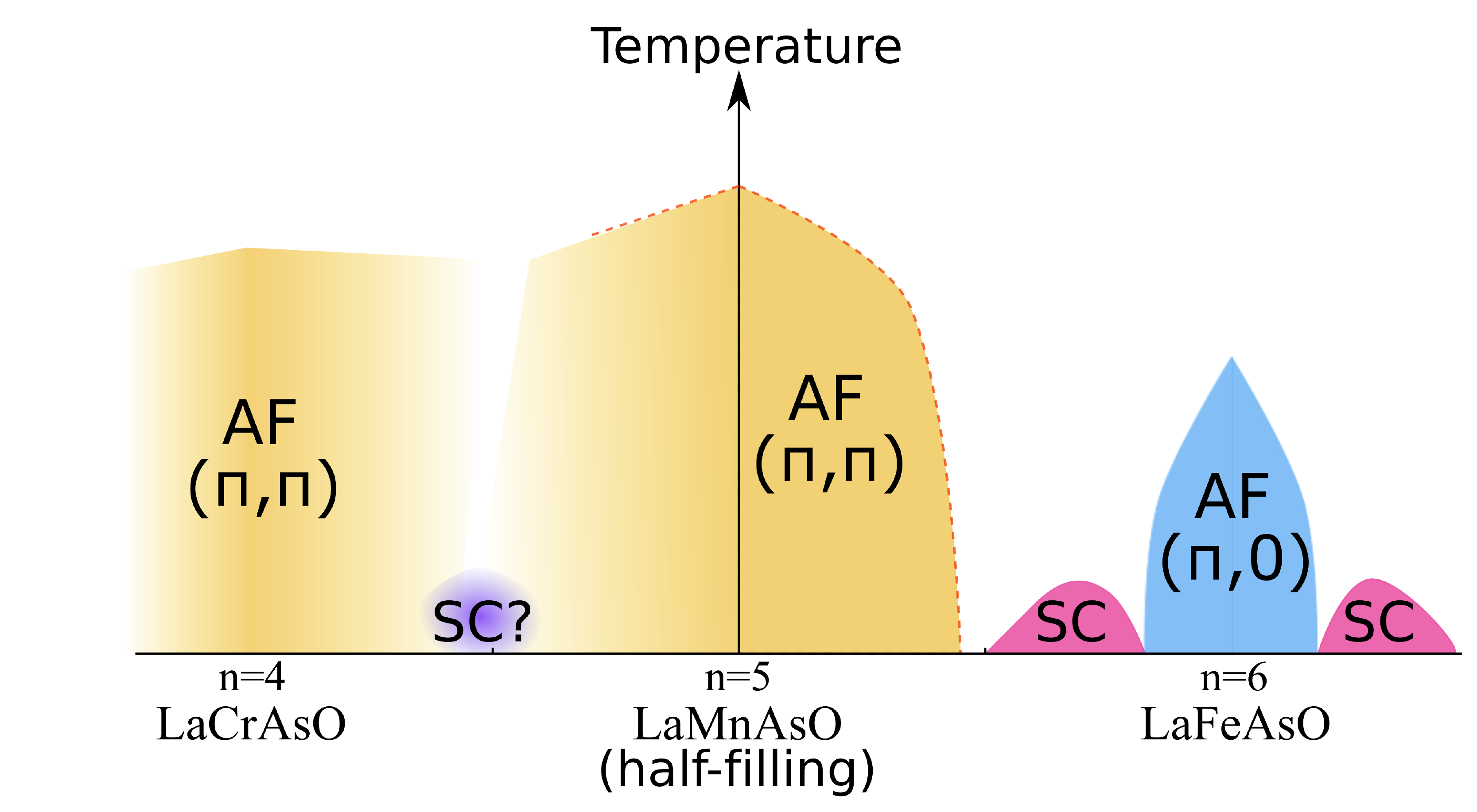} 
\caption{(Color online) Proposed phase diagram for the iron pnictides and isostructural compounds as a function of electronic filling. The half-filled $n=5$ compounds are the Mn-based pnictides which are antiferromagnetic insulators and not superconductors. The electron-doped area around $n=6$ corresponds to the Fe pnictides and chalcogenides, with an antiferromagnetic and metallic region in the center and two superconducting domes on the sides. The hole-doped area around $n=4$ corresponds to Cr-based compounds. From our analysis  and the reference of cuprates we propose that a superconducting dome may arise upon doping Cr compounds if magnetism can be suppressed.} 
\label{fig:Fig0} 
\end{figure}

In 2010, Ishida and Liebsch~\cite{liebsch2010} found theoretically that the correlations in iron superconductors become stronger when the system is doped with holes towards the half-filled Mott-insulating $n=5$ limit. Based on this observation, they proposed a connection between the physics of cuprates and iron based superconductors~\cite{liebsch2010}. In this {\it doped-Mott scenario} the parent compound of iron superconductors is the $n=5$ half-filled system while the usually called undoped materials with $n=6$  are electron-doped systems. Since their proposal, a lot of experimental~\cite{terashima2013,hardy-prl2013,nakajima-scirep014,eilers-prl2016,hardy2016,nosotras_review2016} and theoretical~\cite{liebsch2010b,werner2012,nosotrasprb2012-2,imadaPRL2012,demedici_prl2014,nosotrasprb14,nosotras_review2016} evidence has confirmed the enhancement of correlations when the $n=6$ materials are doped with holes towards half-filling, and their suppression with electron doping away from half-filling.

In iron compounds superconductivity is found for fillings between $n=5.5$ and $6.5$ which means an average doping of $0.1$ to $0.3$ electrons per orbital from the half-filled Mott insulator. These doping values compare well with those for which superconductivity is found in cuprates. The similarities in the correlation dependence and in the doping range at which superconductivity appears in both cuprates and iron superconductors suggest that besides the antiferromagnetic fluctuations an optimum correlation strength is beneficial for achieving high-Tc superconductivity. 

In the doped-Mott scenario the pnictides with layers based on Ni, Pd or Pt, which accomodate $n=8$ electrons, and those with Ir or Rh, with $n=7$, are heavily electron doped and therefore weakly correlated. The insulating $n=5$ Mn-based materials play the role of the Mott insulating parent compounds for which the correlations are very strong, see Fig.~\ref{fig:Fig0}. Taking the cuprates phase diagram as a reference, none of these compounds seems a good candidate for high-Tc superconductivity.  

In this work we  show that for $n<5$ it is possible to find systems with similar correlations  to those found in iron superconductors and propose to search for superconductivity in Cr-based pnictides and chalcogenides.  Cr compounds with $n<5$ electrons (less than one electron per orbital) fit in the range of doping at which the highest critical temperatures are found in hole-doped cuprates. We analyse the properties of these compounds using LaCrAsO as a starting point. We find that when these materiales are doped with electrons it presents mass enhancement factors of the order of those found in the iron pnictides. It also shows competing magnetic tendencies, the most plausible ordering being the $(\pi,\pi)$ checkerboard antiferromagnetism found in cuprates. Finally, its Fermi surface topology favors d-wave superconductivity.

{\it Methods}.
The variety of techniques that we use are explained in detail in the Supplementary Information (SI). We start from a 5-orbital model with on-site interactions: intraorbital $U$ and inter-orbital $U'$ interactions, Hund's coupling $J_H$ and pair-hopping $J'$. We take $U'=U-2 J_H$ and $J'=J_H$ and assume $J_H=0.25U$.

The tight-binding models are formulated within the Slater-Koster approach~\cite{slater54} as detailed elsewhere~\cite{nosotrasprb09}. Parameters are chosen to approximate either the electronic structure of LaFeAsO or LaCrAsO. The LaCrAsO bands are the result of Density functional (DFT) calculations performed using the projector augmented wave method as implemented in the Vienna Ab initio Simulation Package (VASP)~\cite{Kresse_prb1993,Kresse_prb1996,Kresse_prb1999}. The generalized gradient approximation (GGA)  Perdew-Burke-Ernzerhof (PBE) functional was used for the exchange and correlation potentials~\cite{Perdew_prl1996,Perdew_prl1997}.  The crystal structure of the P4/nmm LaCrAsO was fixed to the experimental values~\cite{park-inchem2013}.  The DFT band structure is given in the 2 Cr Brillouin zone. On the other hand, the Fermi surfaces from the tight-binding models are in the unfolded 1 Cr Brillouin zone, whose symmetry points are labelled with 1Cr subscripts. The orbital character is always expressed following the convention of the 1 Cr Brillouin zone with $x$ and $y$ directions along Cr-Cr bonds.

\begin{figure}
\leavevmode
\includegraphics[clip,width=0.23\textwidth]{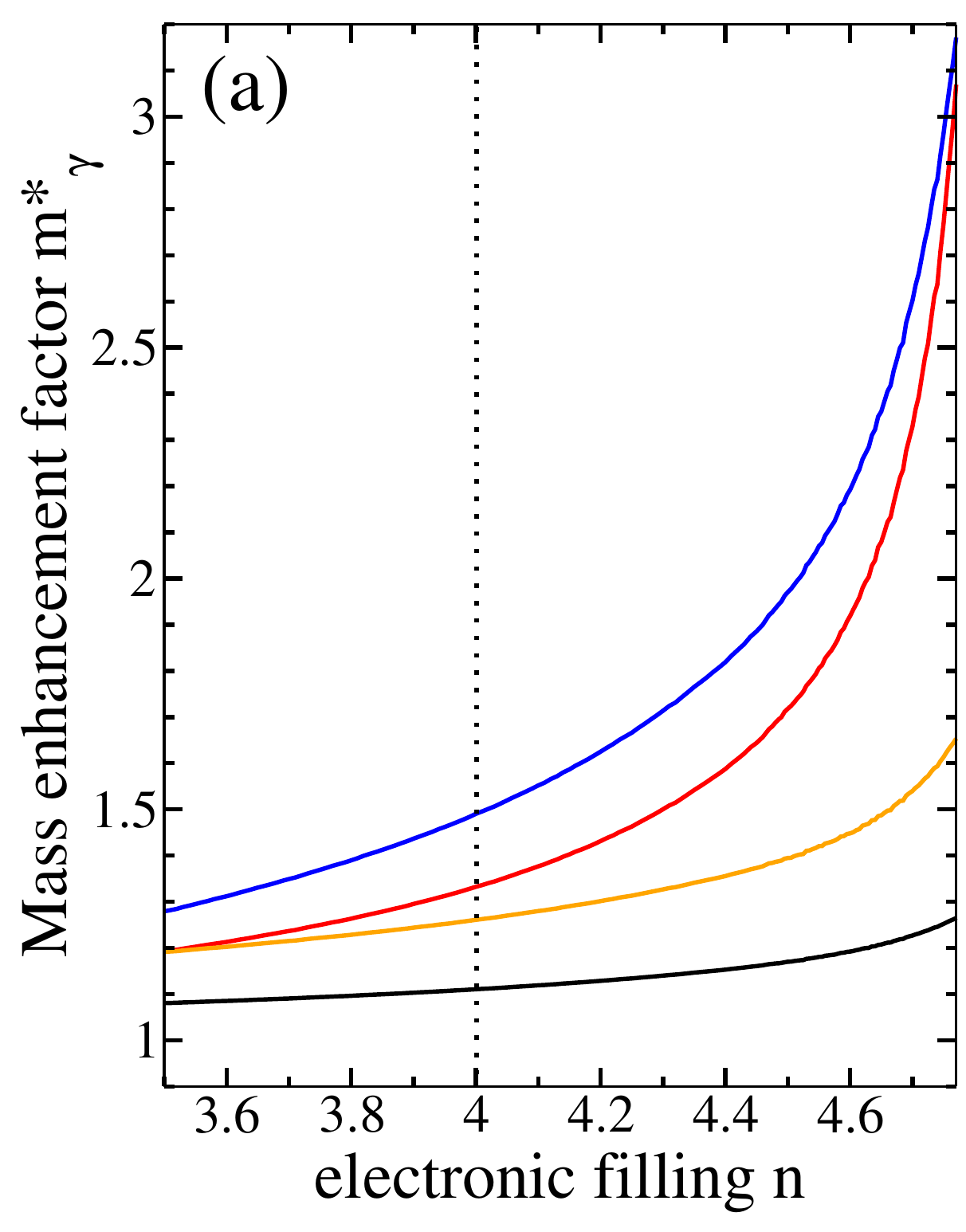} \includegraphics[clip,width=0.23\textwidth]{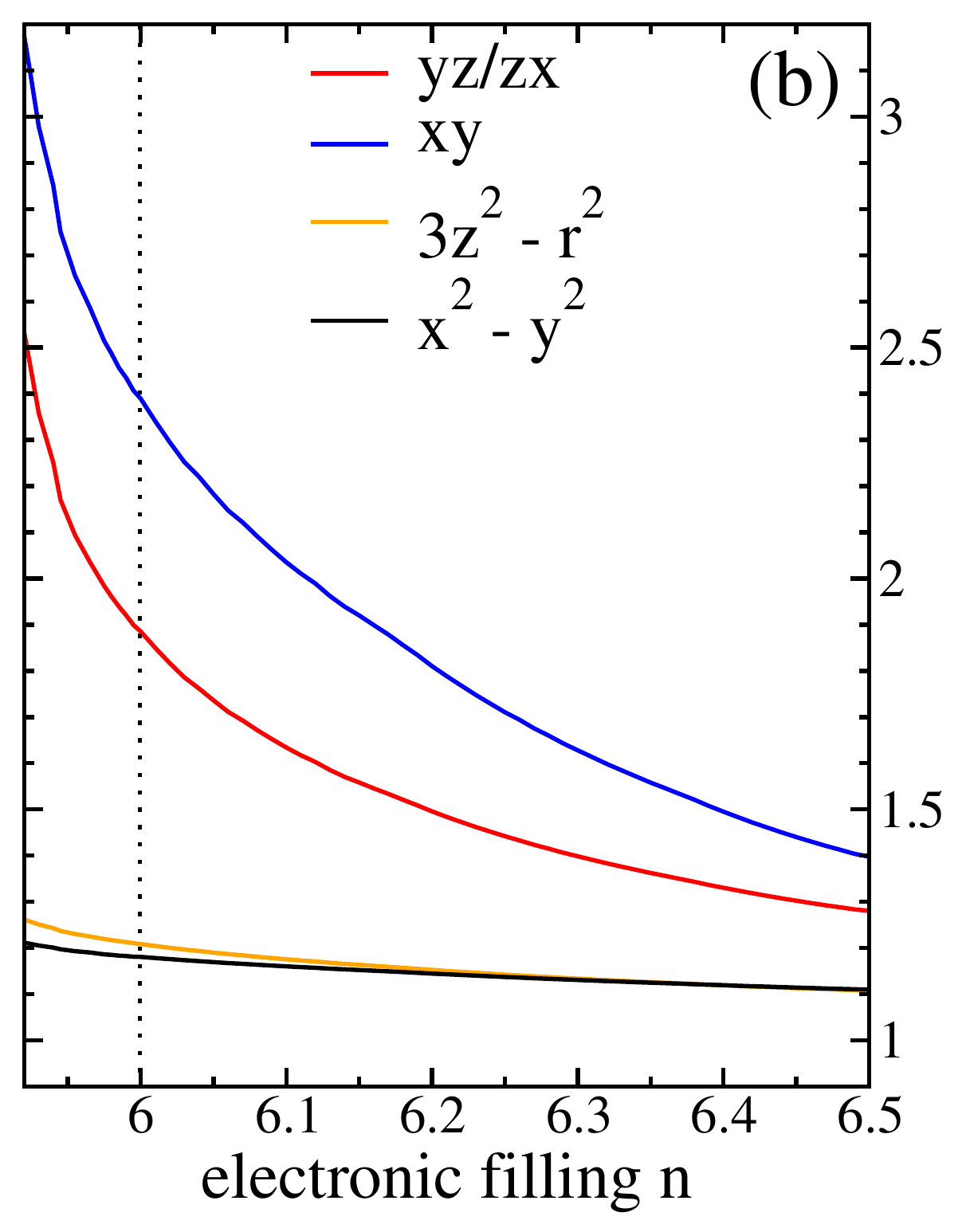} \includegraphics[clip,width=0.23\textwidth]{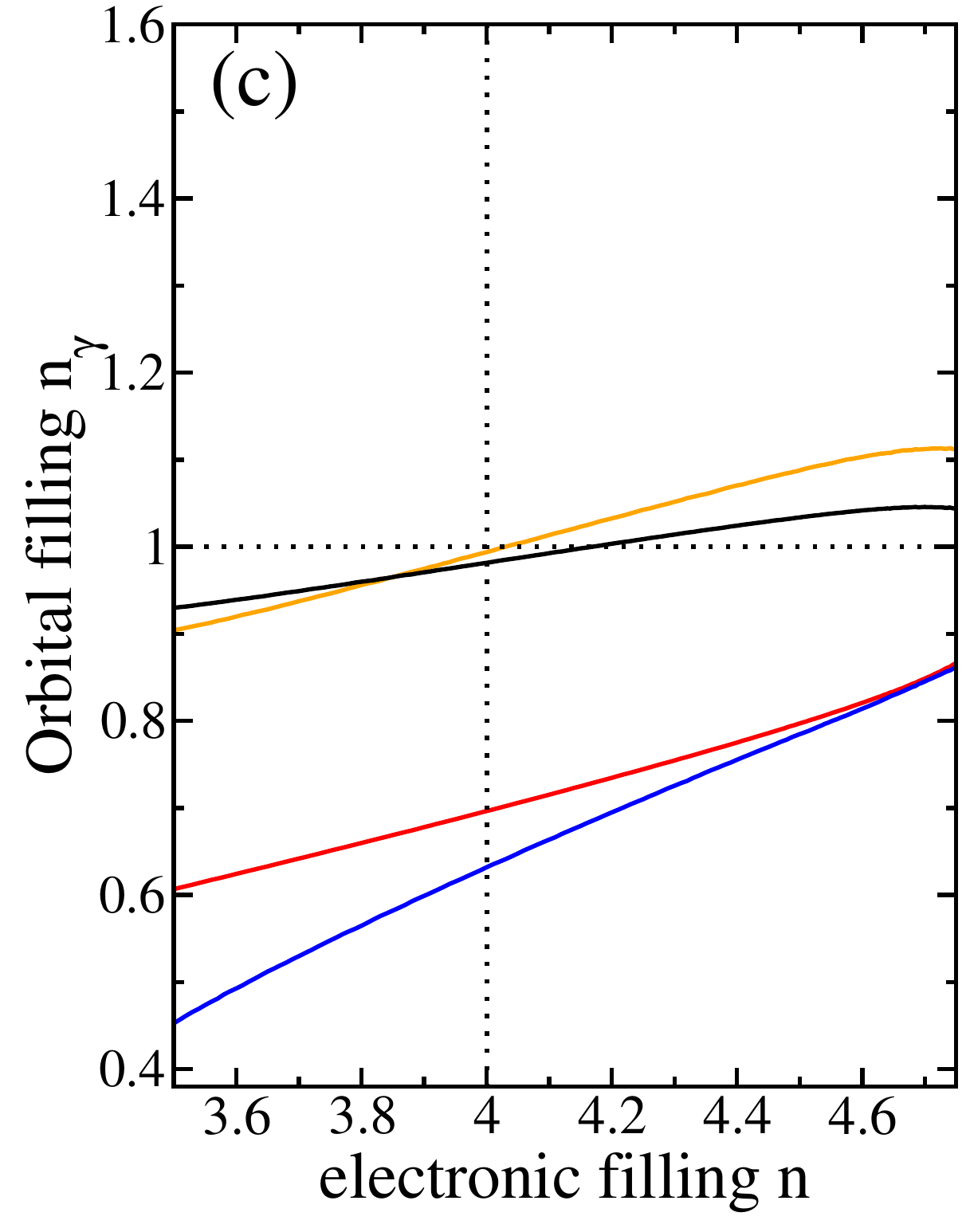} 
\includegraphics[clip,width=0.23\textwidth]{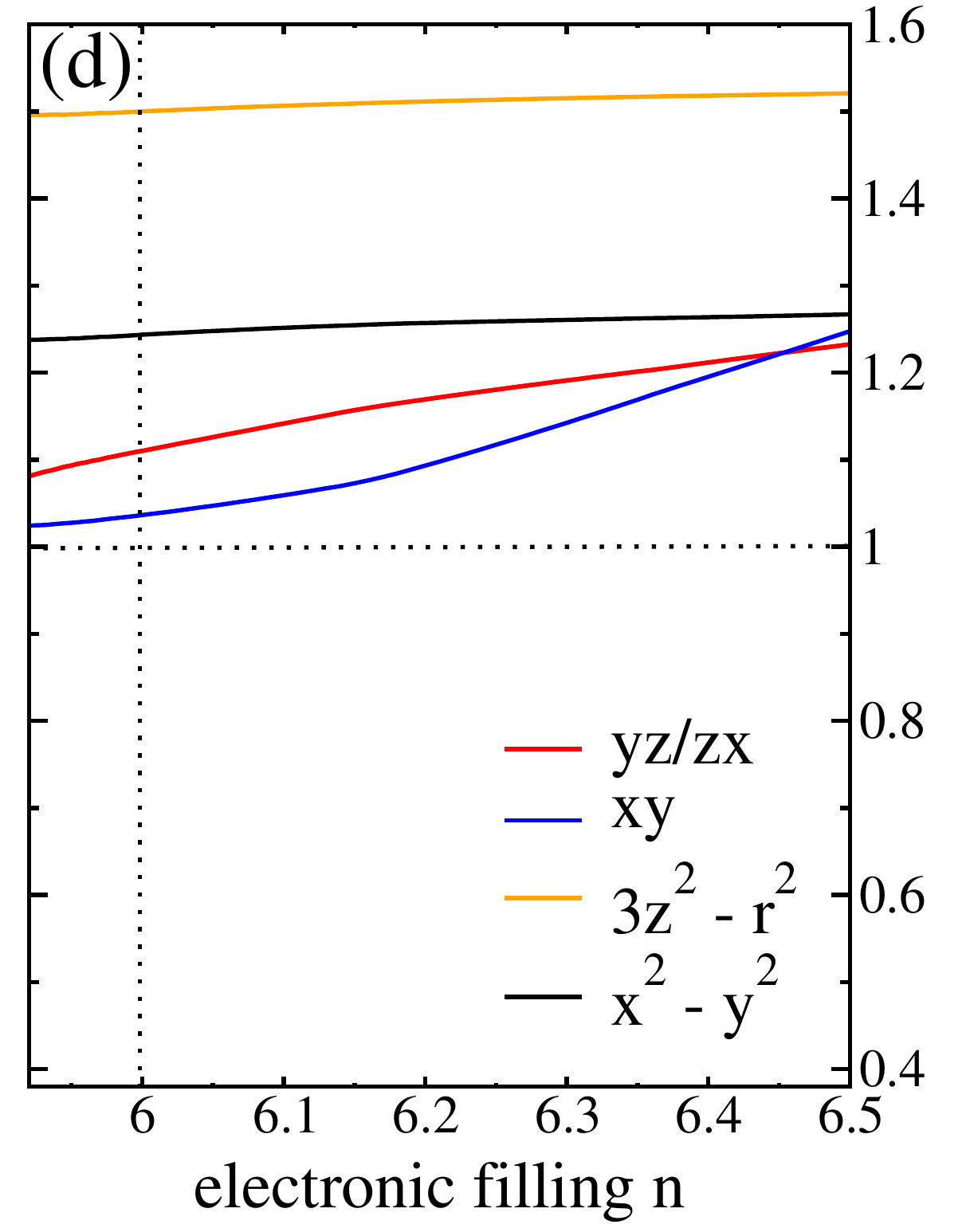} 
\caption{(Color online) (a) and (b) Orbital dependent mass enhancement factors as a function of electronic filling using a tight-binding model proposed for LaFeAsO in Ref.~\cite{nosotrasprb09} and $U_{SL}=3$ eV, see SI. (c) and (d) Orbital filling for the same parameters in (a) and (b). Red is for $yz$ and $zx$, blue for $xy$, yellow for $3z^2-r^2$ and black for $x^2-y^2$.} 
\label{fig:Fig1} 
\end{figure}

We quantify the strength of correlations by the mass enhancement factors $m^*_\gamma$, namely, the ratio between the band mass in the presence of interactions and the one predicted by the DFT calculations. These orbital-dependent mass enhancement factors are given as the inverse of the quasiparticle weights calculated with a slave spin technique~\cite{demedici_prb2005,demedici_prb2010}.
For the analysis of the spin susceptibility and the superconducting order parameter we use  the RPA multiorbital approach~\cite{takimoto_prb2004,kubo_prb2007,graser_njp2008} and focus on singlet pairing solutions.   
The slave spin and RPA approaches deal differently with interactions, hence $U$  takes different values in each case, respectively labelled as $U_{\rm SL}$ and $U_{\rm RPA}$.

{\it Correlations as a function of electronic filling.}  In order to study the effect of doping on the strength of correlations at both sides of half-filling $n=5$ we focus on a tight-binding model proposed for LaFeAsO~\cite{nosotrasprb09}. The orbital dependent mass enhancements $m^*_\gamma$ of this model as a function of filling are plotted in Fig.~\ref{fig:Fig1}.  We use $U_{SL}=3$ eV which gives values of $m^*_\gamma$ similar  to the ones measured in the undoped LaFeAsO.

The increase in correlation strength when the total filling decreases from $n>6$ towards half-filling $n=5$ (Fig.~\ref{fig:Fig1}(b)) is evidenced by the  rise of $m^*_\gamma$. 
The mass enhancements factors are orbital dependent, being much larger for the t$_{2g}$ orbitals $xy$, $zx$ and $yz$ than for the e$_g$ orbitals $3z^2-r^2$ and $x^2-y^2$. Within the context of pnictides, this phenomenon, the so-called orbital differentiation, has been related to the orbital dependent filling~\cite{nosotrasprb2012-2,demedici_prl2014}: the orbitals closer to half-filling are more correlated than those with more electrons, see Fig.~\ref{fig:Fig1}(d).

On the other side of half-filling $n<5$ the mass enhancement factors $m^*_\gamma$ decrease with hole-doping as the system goes away from $n=5$, see Fig.~\ref{fig:Fig1}(a).  This behavior is consistent with the doped Mott scenario~\cite{liebsch2010}. The correlations at both sides of half-filling are different due to the inequivalency of the orbitals. For $U_{\rm SL}=3$ eV, the orbitals which are closer to half-filling (3$z^2$-$r^2$ and $x^2$-$y^2$) are not the most correlated ones, see Fig.~\ref{fig:Fig1}(c). In fact, $xy$ is the most correlated orbital, in spite of being the one farthest from half-filling. This evidences a more prominent role of the orbital bandwidth than previously anticipated. Nevertheless, for larger interaction $U_{\rm SL}$, the orbitals closer to half-filling are more correlated, see SI.

Besides the orbital filling and the orbital bandwidth, the mass enhancement factors $m^*_\gamma$ depend on the total filling $n$ and on the correlations in the other orbitals. In particular, $3z^2-r^2$ and $x^2-y^2$ become closer to half-filling at $n=4$ but they are more correlated at larger $n$.

These results confirm that for $n<5$ it could be possible to find compounds with correlations as strong as those found in iron superconductors. 
However, so far we have just studied the effect of doping on a LaFeAsO model. In the following we focus on a model for the isostructural compound LaCrAsO in order to account for possible changes, driven by the chemical composition, in the band structure and Fermi surface. This may be relevant for superconductivity as, within some formulations of the spin fluctuation theory, it is very sensitive to the shape of the Fermi surface.

\begin{figure}
\leavevmode
\includegraphics[clip,width=0.55\textwidth]{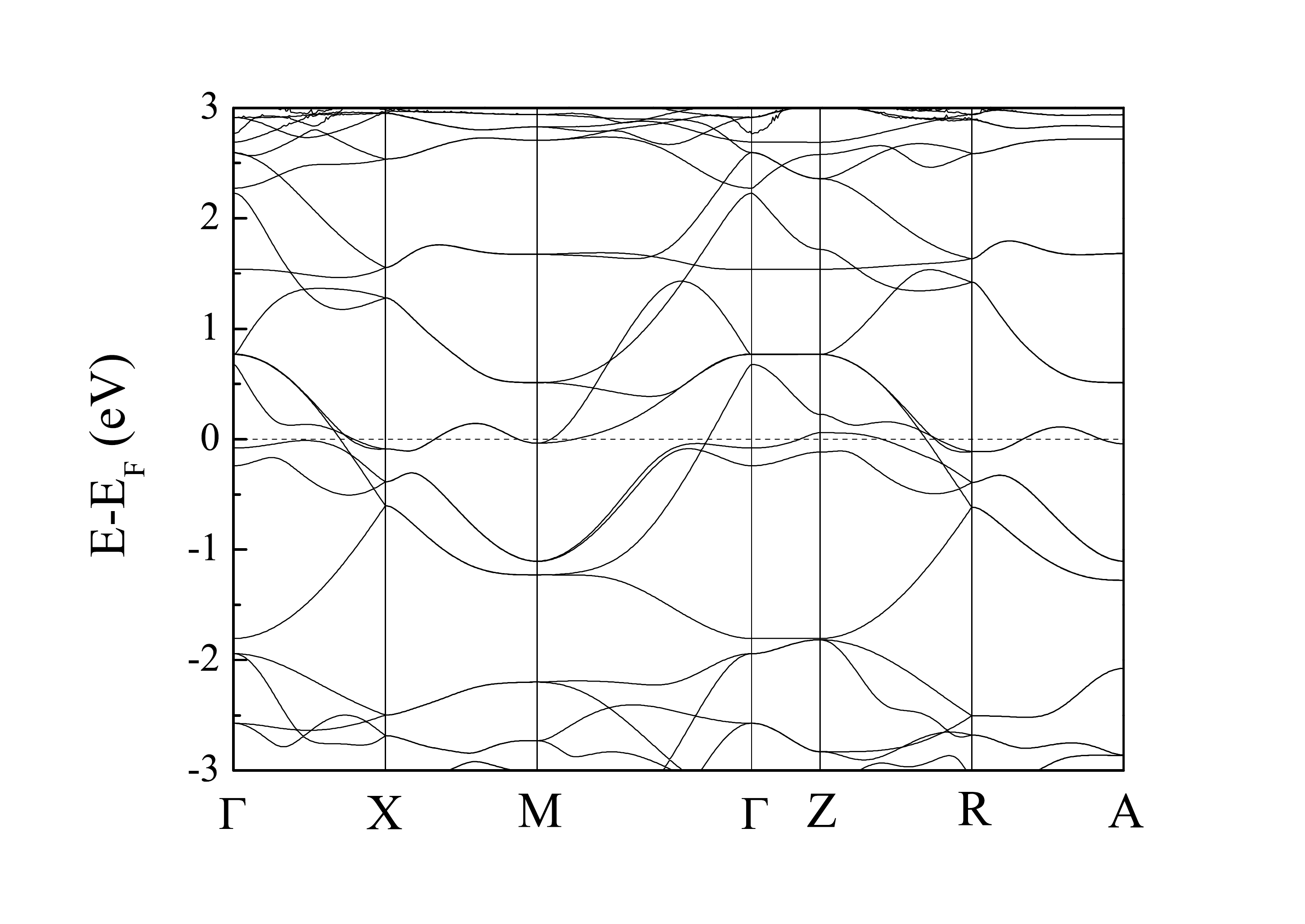}
\vskip -0.5 cm
\caption{ (Color online) DFT electronic structure of LaCrAsO. The bands in the energy range $-2$ to $2$ eV resemble the ones of LaFeAsO. Nevertheless, several bands are shifted with respect to the Fe pnictides affecting the Fermi surface and orbital dependent bandwidths. The Fermi surface of strong two-dimensional character is conformed by a hole-pocket centered at $\Gamma$ and electron pockets at X and M, and symmetry related points. The two bands below the Fermi level along $\Gamma -X$ have $3z^2-r^2$ character. } \label{fig:Fig2}
\end{figure}

{\it LaCrAsO}. {\it Electronic structure}.
Fig.~\ref{fig:Fig2} shows the DFT electronic band structure of LaCrAsO. Bands  between $- 2$ and $2.5$ eV are mostly contributed by Cr atoms. Most of these Cr-bands show quasi-2d behavior. The band structure in this range of energies have similarities with both the LaFeAsO~\cite{vildosola_prb2008} and LaMnAsO bands~\cite{zingl_prb2016} once the shift in chemical potential is taken into account.  On spite of an overall similarity with the electronic structure of these related compounds, differences in the bands close to the Fermi level, which influence the topology of the Fermi Surface, can be appreciated.

Except for a hole 3D pocket centered at Z and absent at $\Gamma$, the Fermi Surface is two-dimensional. In the following we assume that the pocket at $Z$ does not play an important role and focus on the $k_z=0$ plane. The Fermi Surface consists of a flower-shaped hole pocket at $\Gamma$ and shallow electron pockets at X/Y and M in the 2-Cr Brillouin zone (see SI).  With electron doping the size of the electron pockets increases. At $n=4.5$ there is a Lifshitz transition at which the X and M  pockets merge. For larger dopings the Fermi surface consists of three hole pockets centered at $\Gamma$.  

\begin{figure}
\leavevmode
\includegraphics[clip,width=0.23\textwidth]{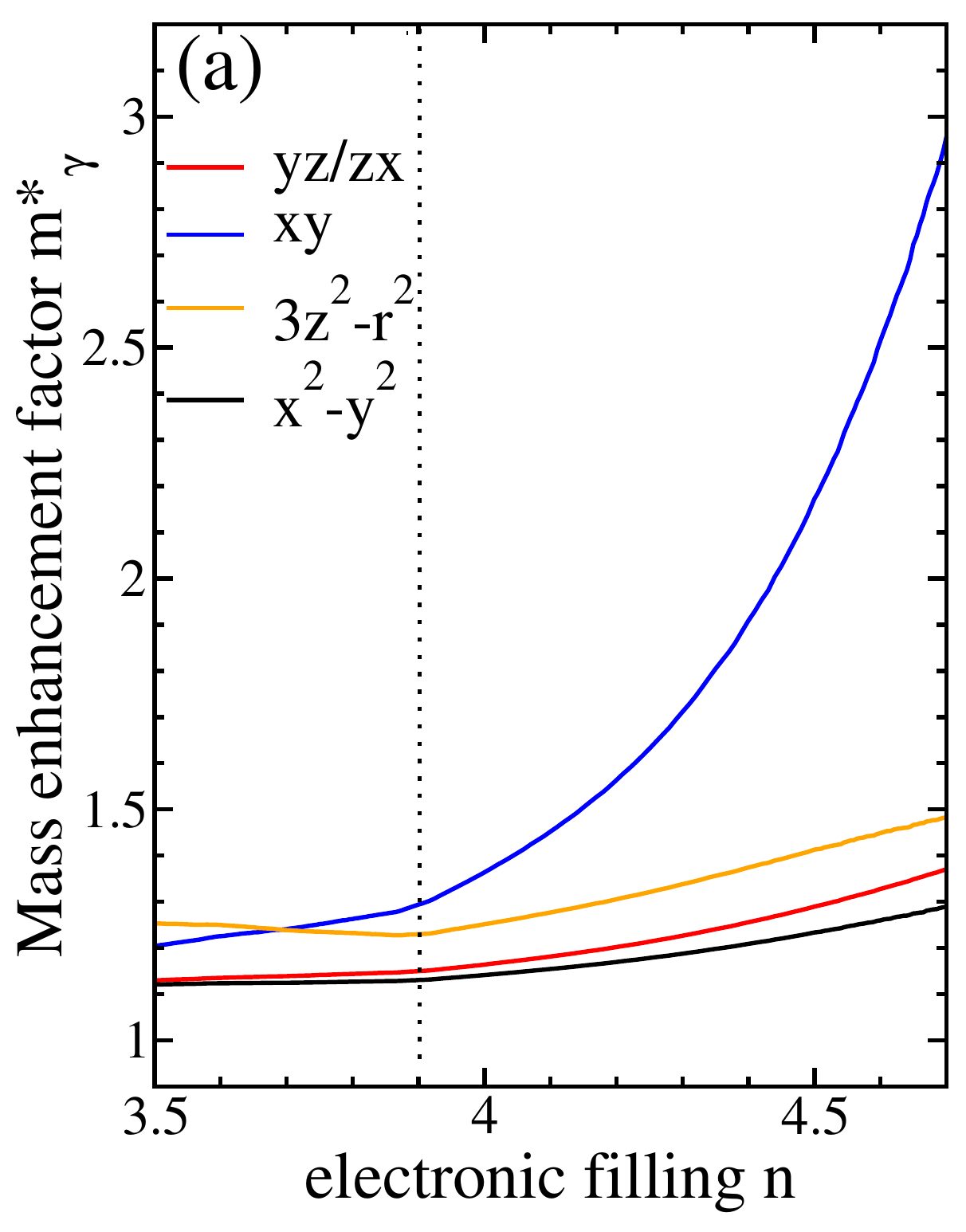}
\includegraphics[clip,width=0.23\textwidth]{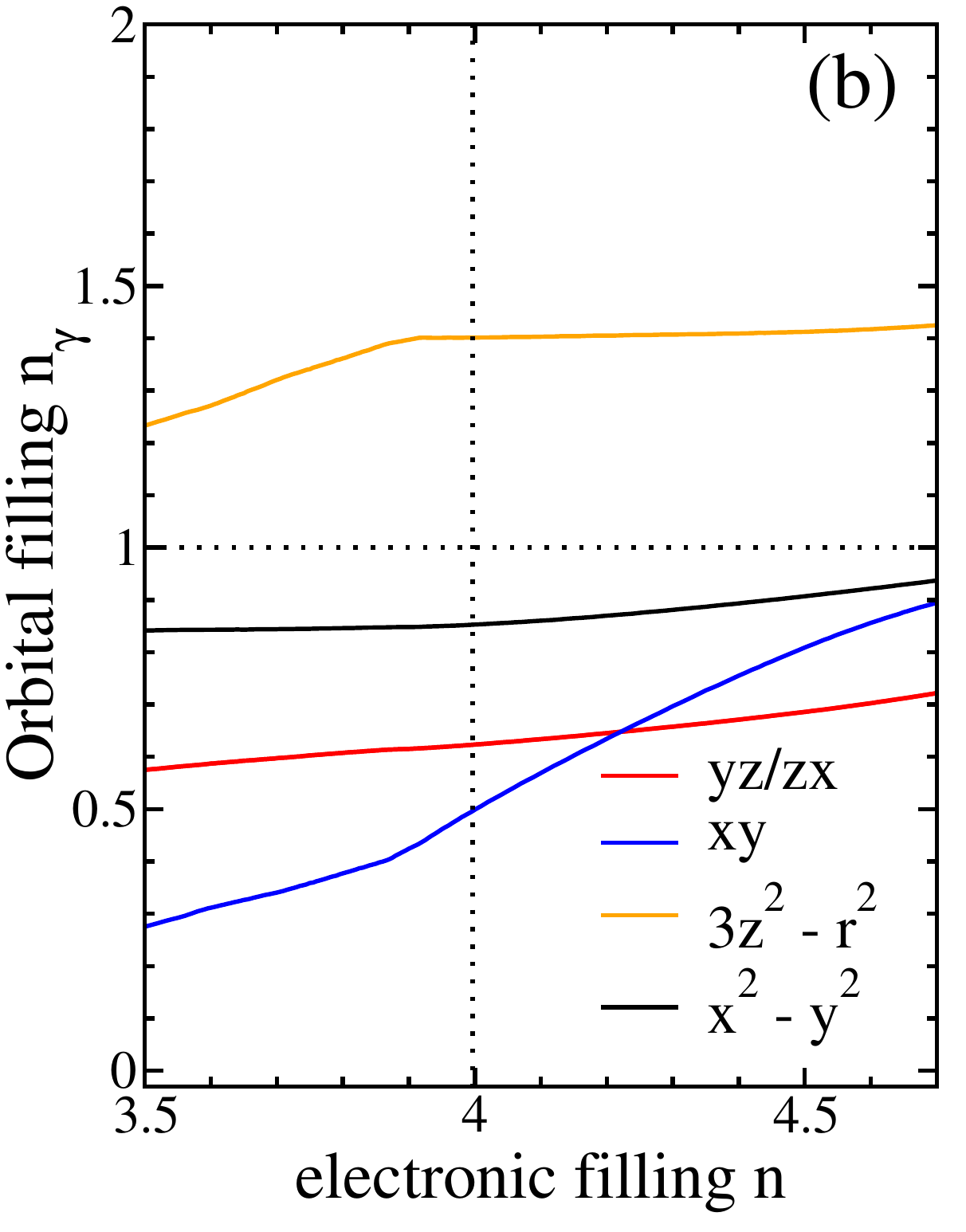}
\caption{ (Color online) Orbital dependent mass enhancement factors $m^*_\alpha$ and orbital fillings $n_\alpha$, respectively in (a) and (b) as a function of the electronic filling using a two-dimensional tight-binding model developed to approximate the bands and the $k_z=0$ Fermi surface of LaCrAsO (see SI). We use $U_{SL}=3$ eV, the same interaction value as in Fig.~\ref{fig:Fig1} for LaFeAsO. } \label{fig:Fig3} 
\end{figure}  

{\it Electronic correlations.}
In agreement with the results presented in Fig.~\ref{fig:Fig1} when LaCrAsO, with nominal filling $n=4$, is doped with electrons towards half-filling ($4<n<5$), the electronic correlations and the mass enhancement factors increase, while they decrease when the system is doped with holes, see Fig.~\ref{fig:Fig3}(a). To our knowledge,
calculations or experiments which address the correlations
in LaCrAsO are not yet available, therefore we have 
adopted for $U_{\rm SL}$ the same value used for LaFeAsO.

Some differences can be appreciated between the mass enhancement factors $m^*_\gamma$ in Fig.~\ref{fig:Fig3}(a) and Fig.~\ref{fig:Fig1}(a).
These differences are linked to the  orbital reorganization of the filling and to the orbital bandwidths. 
In LaCrAsO (with $n=4$ electrons in the Cr d-orbitals)  $n_{3z^2-r^2}\sim1.5$ while $n_{xy}\sim0.5$ electrons (Fig.~\ref{fig:Fig3}(b)). For the same total filling, $n_{3z^2-r^2}\sim1$ in the model for LaFeAsO. The different filling reorganizations can be traced back to the position of the $3z^2-r^2$ bands along $ \Gamma-X$ , which are  below the Fermi level in the DFT electronic structure for LaCrAsO. The distance of the orbitals to half-filling makes them less sensitive to interactions. On the other hand, the bandwidth of the $xy$ orbital is reduced in this compound, which is the most correlated orbital on spite of being far from half-filling in most part of the doping range.

\begin{figure}
\includegraphics[clip,width=0.21\textwidth]{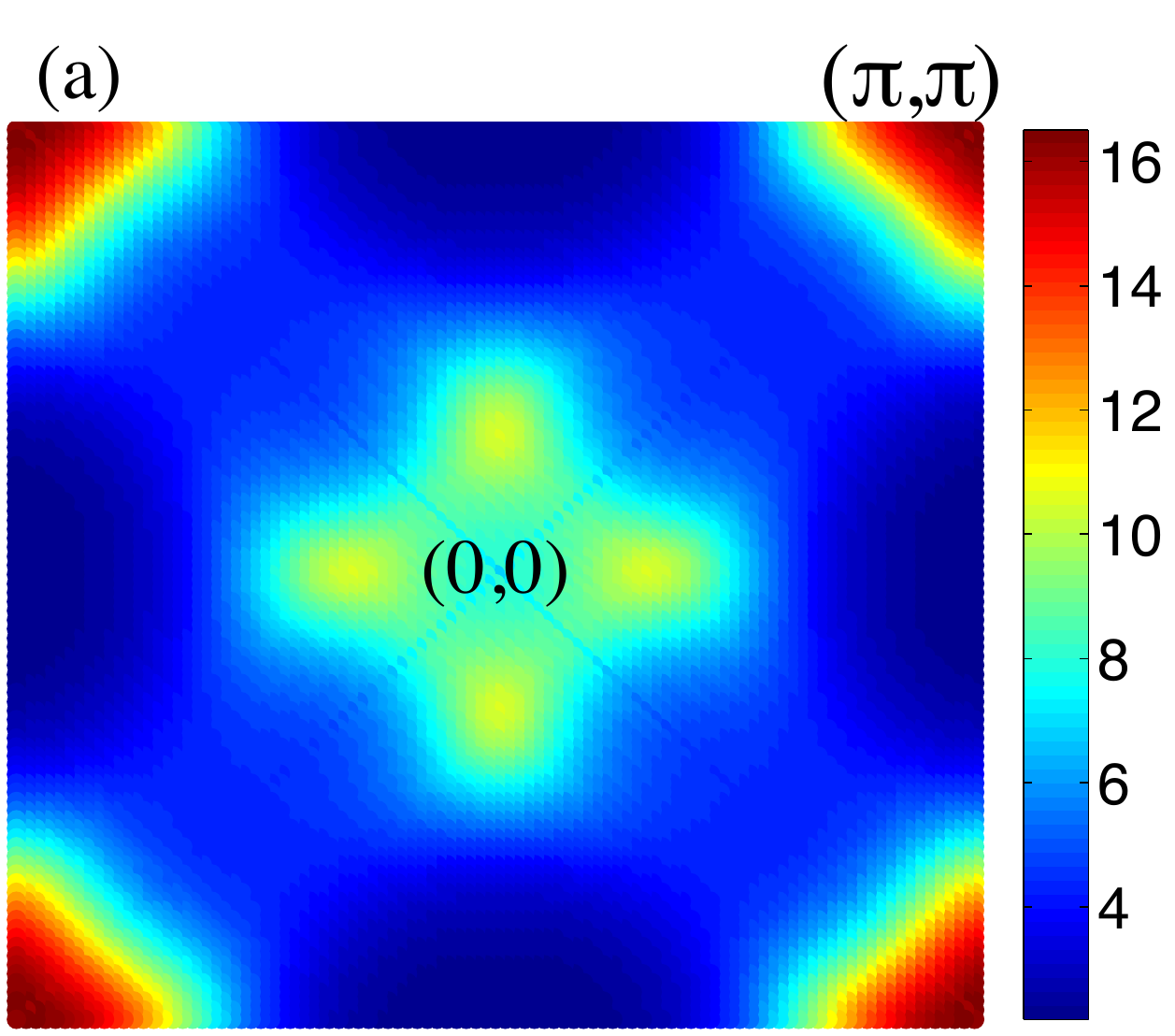}
\includegraphics[clip,width=0.25\textwidth]{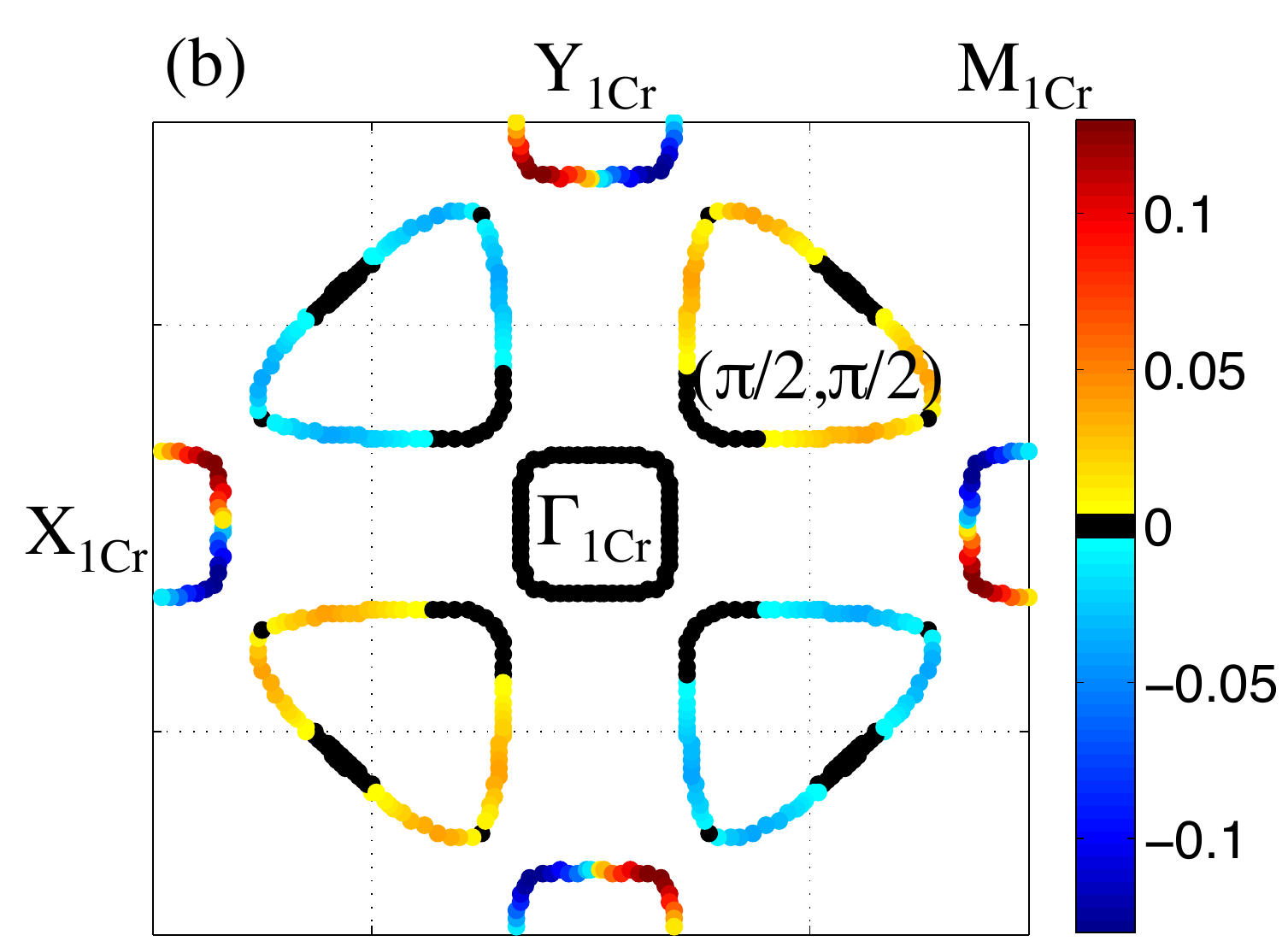}
\caption{(Color online) (a) Spin susceptibility $\chi^{\rm RPA}_{\rm spin}$ in arbitrary units corresponding to electron-doped LaCrAsO using the renormalized electronic structure at $n=4.5$ and $U_{\rm RPA}=0.3$ eV. The susceptibility peaks at $\bf Q=(\pm \pi,\pm \pi)$ and at a slightly inconmensurate momentum $\bf Q_2$ close to $(\pi/2,0)$ or $(0,\pi/2)$. The height of the peaks at $\bf Q $ diverge at $U_c=0.33$ eV. (b) Momentum dependence $g(k)$ of the superconducting order parameter corresponding to the largest eigenvalue of the pairing equation. The d$_{xy}$ symmetry of the order parameter is dictated by fluctuations with momentum $\bf Q_2$.} \label{fig:Fig4} 
\end{figure}  

{\it Magnetism and superconductivity.}
We focus on the electron-doped compounds for which the correlations are sizable and analyse the magnetic tendencies and the most favourable superconducting gap symmetry within RPA.  Here we consider the electronic structure renormalized by the interactions via the mass enhancement factors shown in Fig.~\ref{fig:Fig3}(a),  as we expect the renormalized bands to be a better approximation for the experimental ones. 
In the renormalized band structure the hole pocket at $\Gamma$ becomes slightly smaller and acquires a square like shape. The electron pockets at X become larger and the ones at M smaller. The latter are not present at $n=4$ but re-appear with electron doping. The Lifshitz transition at which the electron pockets merge is found at $n\sim 4.6$.

In the 1Cr-Brillouin zone that we use in Fig.~\ref{fig:Fig4} the hole-pocket is centered at $\Gamma_{1Cr}$ and the electron pockets are respectively found around $(\pm \pi/2,\pm \pi/2)$, $Y_{\rm 1Cr}=(0,\pm \pi)$, and $X_{\rm 1Cr}=(\pm \pi,0)$ for fillings below the Lifshitz transition. Above this transition, the Fermi surface consists of two hole pockets (one large and one small) centered at $\Gamma_{\rm 1Cr}$ and a large hole pocket centered at $M_{\rm 1Cr}$. Nesting features appear both below and above the transition. 

Within the range $n \sim 4.4-4.7$ and for interactions below the magnetic instability, the spin susceptibility $\chi^{\rm RPA}_{\rm spin}$ is 
enhanced around $\bf Q=(\pi,\pi)$ and around a slightly incommensurate vector $\bf Q_2$ close to $(\pi/2,0)$ or $(0,\pi/2)$. $\chi^{\rm RPA}_{\rm spin}$ is plotted in Fig.~\ref{fig:Fig4} for $n=4.5$ and $U_{\rm RPA}=0.3$ eV.   The two peak structure of  $\chi^{\rm RPA}_{\rm spin}$  reveals the presence of competing antiferromagnetic instabilities. The relative height of the peaks is doping and interaction dependent.  With increasing interaction the peak at $(\pi,\pi)$ diverges at $U_c$, indicating a transition to a checkerboard antiferromagnetic ground state~\cite{fsnonren}. 

We now focus on the leading superconducting instability in this range of dopings at interactions below $U_c$. We find that the largest eigenvalue of the pairing equation corresponds to an order parameter $\Delta g(k)$ with d$_{xy}$ symmetry except maybe for interactions very close to $U_c$. The order parameter, whose momentum dependence $g(k)$ is plotted in Fig.~\ref{fig:Fig4}(b) for $n=4.5$, changes sign at the $x$ and $y$ axis.
The maximum amplitude of the order parameter is found along the electron pockets at $X_{1\rm Cr}$ and $Y_{1\rm Cr}$. Interestingly the leading superconducting symmetry is not dictated by the momentum $\bf Q$ at which the spin susceptibility diverges, but by $\bf Q_2$.  The enhanced response at $\bf Q_2$ originates in the scattering between the tips of the electron pockets. This scattering is strongly enhanced due to the proximity of the Lifshitz transition.

{\it Discussion.}
We find strong correlations and antiferromagnetic tendencies for $n<5$ that could provide a breeding ground for a yet unobserved superconductivity phase in chromium compounds. 



Strong correlations decreasing with hole-doping are expected in different Cr pnictides and chalcogenides with $4<n<5$. The values of the orbital dependent correlations strength may be influenced by details of the specific band structure. In particular, the position of the $3z^2-r^2$ and $xy$ bands, which appear close to the Fermi level or crossing it, is very sensitive to small changes in the lattice parameters~\cite{vildosola_prb2008,nosotrasprb09}.

The momentum dependence of the spin susceptibility of electron doped LaCrAsO suggests the presence of competing antiferromagnetic tendencies. $(\pi,\pi)$ is not only the most plausible ordering within the RPA weak coupling approach but it has also been found in Hartree-Fock calculations (not shown) at larger values of the interaction. Experimentally, this ordering is present in both LaCrAsO ($n=4$) and {\mbox LaMnAsO} ($n=5$)~\cite{emery_prb2011,park-inchem2013,mcguire_prb2016}. 
Hence, it would not be surprising if antiferromagnetic ordering is also found in the chromium electron-doped compounds.  In such case, to observe the hypothetic superconducting phase predicted here, the magnetic phase should be suppressed with pressure or chemical substitutions. 

In our calculations for LaCrAsO the d$_{xy}$ symmetry of the superconducting instability is determined by the proximity of a Lifshitz transition between the electron pockets at $(\pi,0)$ and $(\pi/2,\pi/2)$ (in 1Cr Brillouin zone). Both electron pockets are very shallow and could be absent in other related compounds. In such a case a different pairing symmetry could be preferred. 

Various Cr pnictides have already been synthesized with the 1111, 122 and 2322 structures~\cite{singh-prb2009,park-inchem2013,paramanik-prb2014,jiang-prb2015}. 
To our knowledge, CrSe has not yet been synthesized in the PbO-structure but, according to ab-initio calculations, it is expected to be stable~\cite{ding_ssc2009}. In these and other families of pnictides and chalcogenides, our proposal opens a new avenue to search for unconventional superconductivity.

We thank conversations with L. Brey, L. Fanfarillo and B. Valenzuela. 
Funding from Ministerio de Econom\'ia y Competitividad via grants No. FIS2012-33521, FIS2014-53218-P, FIS2015-64654-P, MAT2015-66888-C3-1-R.
and from Fundaci\'on Ram\'on Areces is gratefully acknowledged. J.L. acknowledges the financial support of the Natural Science Foundation of China (11374186 and 51231007) and the China Scholarship Council.

\beginsupplement
\section{SUPPLEMENTARY INFORMATION: METHODS}
\subsection*{Ab-initio calculations}

Density functional calculations are performed using the projector augmented wave method as implemented in the Vienna Ab initio Simulation Package (VASP)\cite{Kresse_prb1993,Kresse_prb1996,Kresse_prb1999}. The generalized gradient approximation (GGA)  Perdew-Burke-Ernzerhof (PBE) functional was used for the exchange and correlation potentials\cite{Perdew_prl1996,Perdew_prl1997}.  The energy cutoff for the plane waves expansion of the electron wave function was settled to 500 eV and the bulk Brillouin zone is sampled with a $\Gamma$-centered 7 $\times$ 7 $\times$  3 k-point grid for self-consistency and a 15 $\times$ 15 $\times$ 7 grid for the calculation of the density of states. The crystal structure of the P4/nmm LaCrAsO was fixed to the experimental values a = b = 4.0412 Å and c = 8.9863 $\AA$, and La and As were located at 0.1365 and 0.6625 $\times$ c, respectively~\cite{park-inchem2013}.

\begin{figure}
\includegraphics[clip,width=0.44\textwidth]{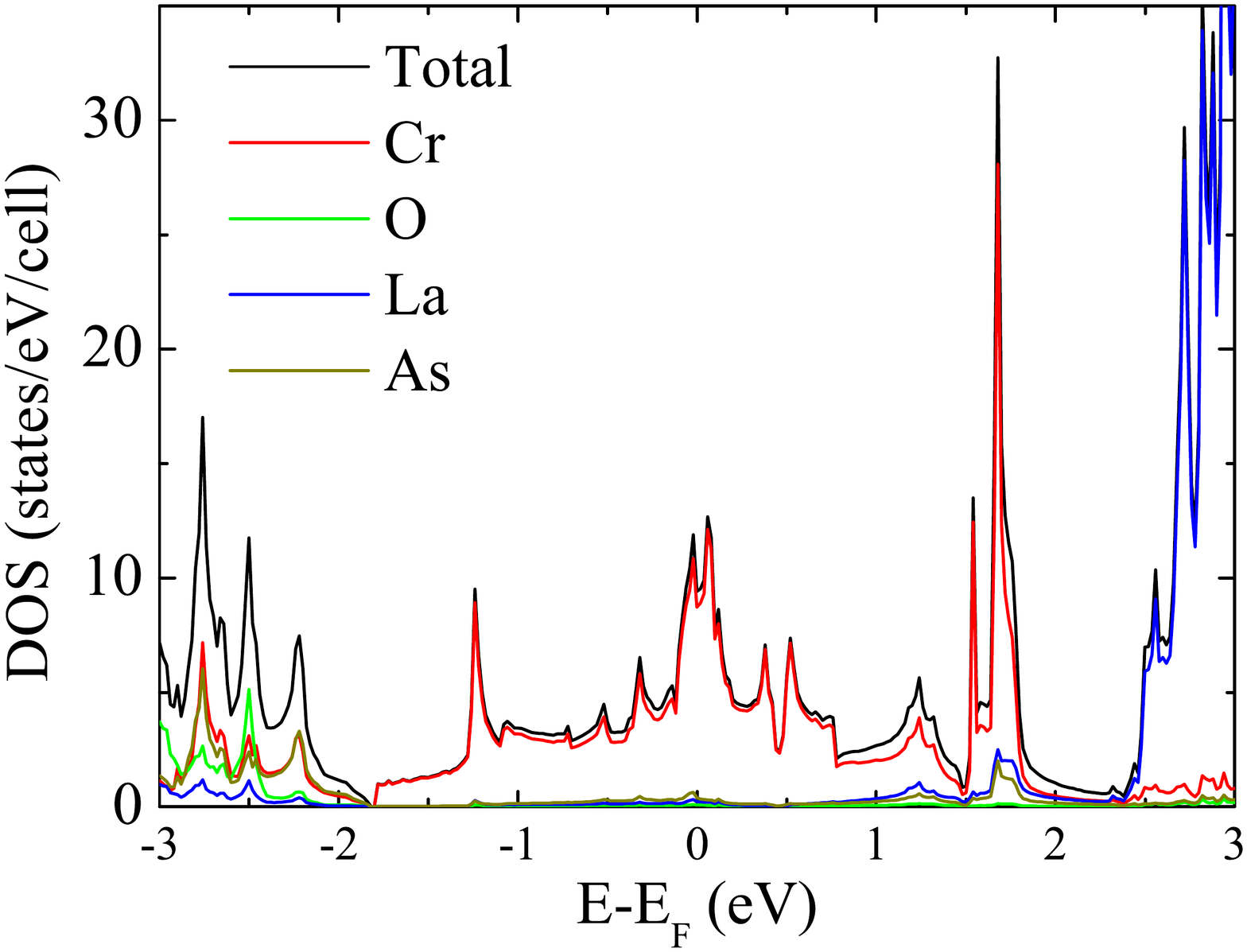} 
\includegraphics[clip,width=0.22\textwidth]{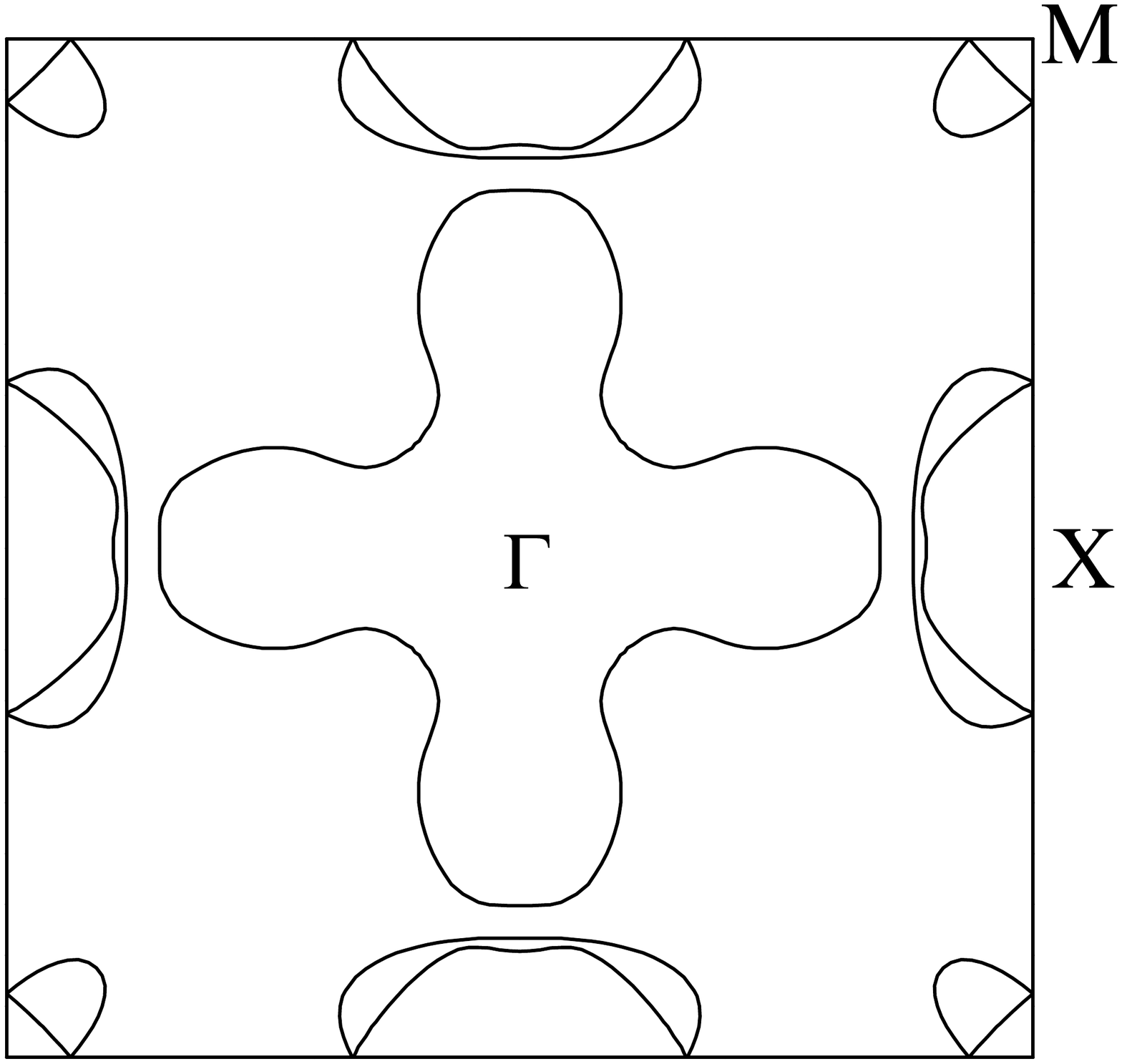}
\includegraphics[clip,width=0.22\textwidth]{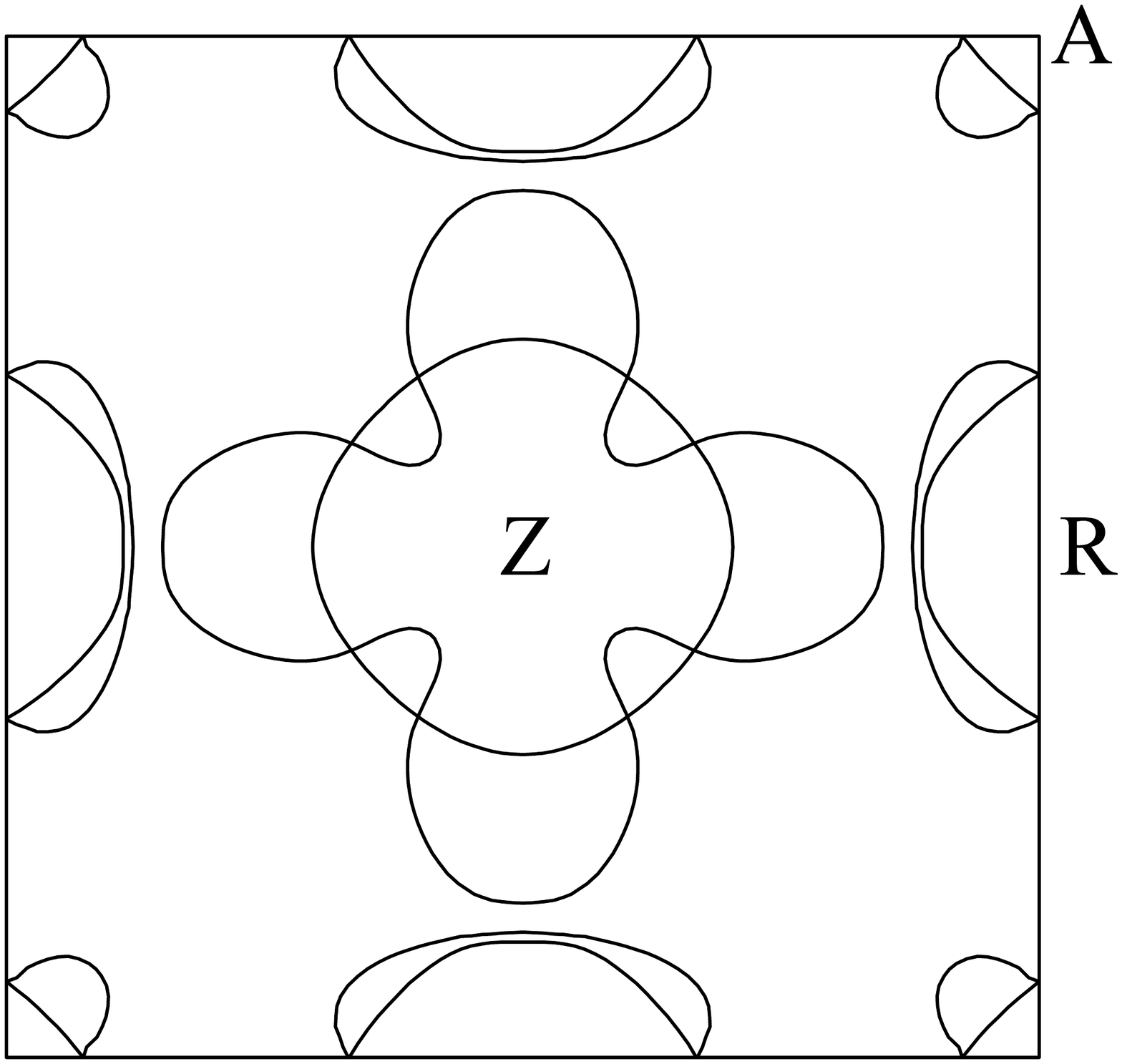}
\caption{(Color online) Top: Total and partial densities of states of each type of electron in LaCrAsO as a function of energy calculated in GGA approximation. The energy spectrum between -2 eV and 2 eV is contributed mostly by Cr d-electrons. Fermi surface at the $k_z=0$ (bottom left) and $k_z=\pi$ (bottom right) planes. }
\label{Fig:fig.S1}
\end{figure}

The electronic band structure is given in Fig.~3 in the main text. As observed in Fig.~\ref{Fig:fig.S1} where the partial density of states is plotted, the bands around the Fermi surface are mostly contributed  by Cr d-electrons, similarly to the Fe d-electrons in the iron pnictides.  In the 2Cr Brillouin zone the Fermi surface consists of a two-dimensional flower shaped hole-pocket at $\Gamma$ and shallow electron pockets at X and M. Moreover, there is a three dimensional large and shallow hole pocket at Z.

\subsection{The model}
To study the electronic properties we start from a five-orbital model with local interactions including: intraorbital $U$, interorbital $U'$, 
Hund's coupling $J_H$, and pair hopping $J'$ terms,

\begin{eqnarray}
\nonumber
  H   = \sum_{k,\gamma,\beta,\sigma}\epsilon_{k,\gamma,\beta}c^\dagger_{k,\gamma,\sigma}c_{k,\beta,\sigma}+h.c. + \sum_{j,\gamma,\sigma}\epsilon_\gamma n_{j,\gamma,\sigma}
\\ \nonumber
 +  U\sum_{j,\gamma}n_{j,\gamma,\uparrow}n_{j,\gamma,\downarrow}
 +  (U'-\frac{J_H}{2})\sum_{j,\gamma>\beta,\sigma,\tilde{\sigma}}n_{j,\gamma,\sigma}n_{j,\beta,\tilde{\sigma}}
\\ 
 -  2J_H\sum_{j,\gamma >\beta}\vec{S}_{j,\gamma}\vec{S}_{j,\beta}
 +   J'\sum_{j,\gamma\neq
  \beta}c^\dagger_{j,\gamma,\uparrow}c^\dagger_{j,\gamma,\downarrow}c_{j,\beta,\downarrow}c_{j,\beta,\uparrow}
 \,
\label{eq:hamiltoniano}
\end{eqnarray}
$i,j$ label the Fe/Cr sites in the 1 Fe/Cr unit cell, $k$ the momentum in the 1Fe/Cr Brillouin zone, $\sigma$ 
the spin and $\gamma$, and $\beta$ the
five Fe/Cr d-orbitals $yz$, $zx$, $xy$, $3z^2-r^2$ and $x^2-y^2$, with $x$ and $y$ axis along the Fe-Fe/Cr-Cr bonds. We use
$U'=U-2J_H$~\cite{castellani78} and $J'=J_H$, as in rotationally invariant systems, leaving only two independent interaction parameters, $U$ and $J_H$. We take $J_H=0.25 U$. We use $U_{\rm SL}$ and $U_{\rm RPA}$ to refer to the value of $U$ in the slave spin and the RPA calculations, respectively.

We consider a 2D tight-binding model to mimic the non-renormalized band structure of LaFeAsO and LaCrAsO. Each case is described by a different set of parameters.
To build the tight-binding of the pnictide layers we follow the Slater-Koster method~\cite{slater54} as described in~\cite{nosotrasprb09}.  In this model the hopping parameters are written in terms of the overlap integrals between the Fe/Cr d-orbitals (dd-overlap terms, direct hopping) and between the Fe/Cr d-orbitals and the As p-orbitals (pd-overlaps, indirect hopping).
The hopping amplitudes, restricted to first and second neighbors, depend on the angle $\alpha$ formed by the Fe-As/Cr-As bonds and the Fe/Cr-plane~\cite{nosotrasprb09}. We take $\alpha=35.3^o$, corresponding to the regular Fe-As/Cr-As tetrahedra. In both cases the energies are  in units of $(pd\sigma)^2/|\epsilon_d-\epsilon_p|$ with $|\epsilon_d-\epsilon_p|$ the energy difference between the Fe/Cr-d orbitals and the As-p orbital~\cite{nosotrasprb09}. We take $(pd\sigma)^2/|\epsilon_d-\epsilon_p|= 1$ eV.

To describe the iron pnictide compound we use the overlap and crystal field parameters given in~\cite{nosotrasprb09} and used extensively afterwards to study the electronic properties of iron pnictides. The electronic bands corresponding to this model are given in~\cite{nosotrasprb09} and plotted with colored orbital dependent weights in~\cite{nosotrasprb13}. 

\begin{figure}
\includegraphics[clip, width=0.5\textwidth]{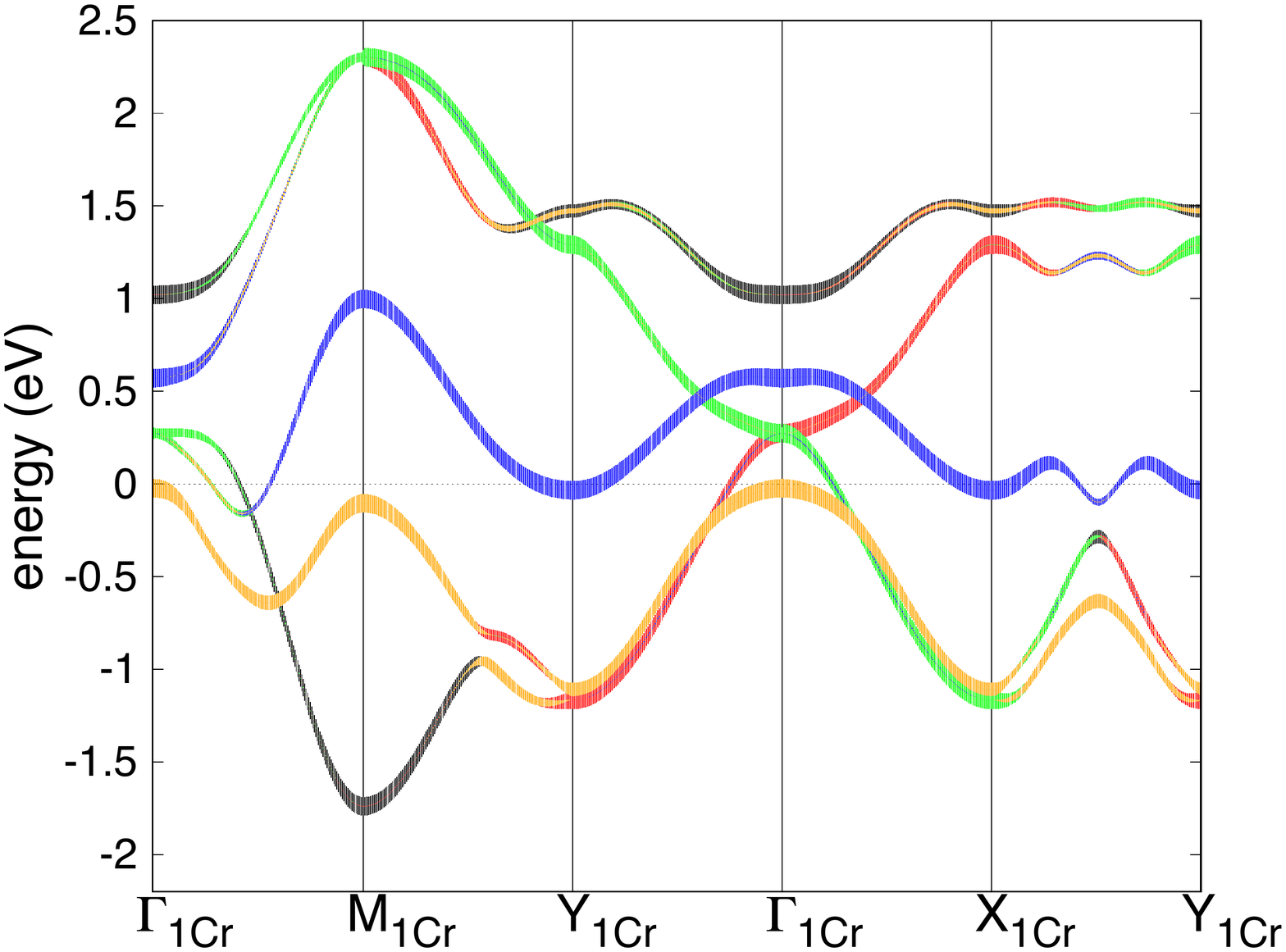}
\caption{Electronic structure of LaCrAsO computed with the approximated tight-binding model introduced in the text. The 1Cr Brillouin zone is used with $x$ and $y$ axis along the Cr-Cr bonds. The linewidth of the bands gives the orbital content. Here $xy$ is given in blue, $yz$ in red, $zx$ in green, $3z^2-r^2$ in yellow and $x^2-y^2$ in black.}
\label{fig:Fig.S2}
\end{figure}

For the Cr compound we take the crystal field parameters 
$\epsilon_{xy}=-0.3$, $\epsilon_{yz,zx}=0$, 
$\epsilon_{3z^2-r^2}=-0.9$ and  $\epsilon_{x^2-y^2}=-0.48$. 
The overlap integrals used are: $pd_\sigma=0.648$, $pd_\pi=-0.456$, 
$dd_{\sigma 1}=-0.42$, $dd_{\pi 1}=0.36$, $dd_{\delta 1}=-0.12$ and
$dd_{\sigma 2}=-0.024$. Moreover we substract an amplitude 0.2 to the first nearest neighbor hoppings $ t^{yz,yz}_{i,i+1}$ and symmetry related ones. The resulting electronic structure is plotted in Fig.~\ref{fig:Fig.S2}. The width of each line gives the orbital weight following the color code: blue $xy$, red $yz$, green $zx$, yellow $3z^2-r^2$ and black $x^2-y^2$. 

\subsection{Techniques for multi-orbital models}

{\it \bf Slave-spin technique.} To analyze the correlation strength we have used the Z$_2$ slave-spin technique developed in~\cite{demedici_prb2005,demedici_prb2010} in its single-site approximation, see also~\cite{fanfarillo_prb2015,demedici2016}. In short, in this slave-spin approach the physical fermions are written in terms of pseudospin operators and auxiliary fermions. The two states of the pseudospin represent the two possible occupancies of a spinless fermion on a given site (0 and 1). The auxiliary fermions are introduced to ensure the anticommutation relations. As in other slave particle approaches this procedure generates unphysical states which are eliminated by imposing a constraint via Lagrange multipliers $\lambda_\gamma$, where $\gamma$ labels the orbital. The constraint is satisfied only at the mean-field level.  This slave spin method has been previously used to study the correlations in iron superconductors~\cite{demedici_prl2014,nosotrasprb14,fanfarillo2016} and in Hund metals~\cite{fanfarillo_prb2015,demedici_prb2011}. In this approach Hund's coupling is treated at the Ising level and the pair-hopping does not enter.

This technique allows to calculate the orbital dependent quasiparticle weight $Z_\gamma$. A quasiparticle weight smaller than unity evidences the presence of correlations and narrows the bands. In the single-site approximation used in this work, the quasiparticle weight is equal to the inverse of the orbital dependent mass enhancement factor $m^*_{\gamma}$, see~\cite{demedici_prb2010} for a discussion. 

When the physical fermion is written in terms of the slave variables an arbitrary gauge parameter $c_\gamma$ is introduced. The value of  $c_\gamma$ has to be fixed by imposing physical conditions. Imposing that the quasiparticle weight $Z_\gamma$ is equal to unity when the interactions vanish $c_\gamma$ becomes equal to $c_\gamma=1/\left(n_{\gamma,\sigma}(1-n_{\gamma,\sigma})^{1/2})\right )-1$ with $n_{\gamma,\sigma}=n_\gamma/2$ the orbital occupation per spin~\cite{demedici_prb2010}. We use this expression for the gauge parameter using the self-consistent orbital filling at a given interaction $U_{\rm SL}$.

The Lagrange multipliers $\lambda_\gamma$  give the orbital-dependent onsite energy shifts induced by the interactions.  To compensate the finite values of the Lagrange multipliers which appear in the Z$_2$ implementation of the slave-spin technique at $U_{\rm SL}=0$, extra Lagrange multipliers $\lambda_{0,\gamma}$ are introduced. $\lambda_{0,\gamma}$ add to the onsite energy terms as $\tilde \epsilon_\gamma = \epsilon_\gamma + \lambda_{0,\gamma}$ in the Hamiltonian of the auxiliary fermions.  The value of $\lambda_{0,\gamma}$ used, kept fixed for all values of $U_{\rm SL}$, is given by 
\begin{equation}
\lambda_{0,\gamma}=-4 \frac{n_{\gamma \sigma}-1/2}{n_{\gamma \sigma}(1-n_{\gamma \sigma})}\left |\sum_\beta \sum_k \epsilon_{k,\gamma,\beta}\langle c^\dagger_{k,\gamma,\sigma} c_{k,\beta,\sigma}\rangle \right|.
\end{equation}
In the non-magnetic state, $\lambda_{0,\gamma}$ and $\langle c^\dagger_{k,\gamma,\sigma} c_{k,\beta,\sigma} \rangle$  do not depend on spin $\sigma$.
Here the  fillings $n_{\gamma\sigma}$ and the occupations $\langle c^\dagger_{k,\gamma,\sigma} c_{k,\beta,\sigma} \rangle$ are calculated using the non-interacting part in Eq.~(\ref{eq:hamiltoniano}). 

\begin{figure}
\includegraphics[clip, width=0.5\textwidth]{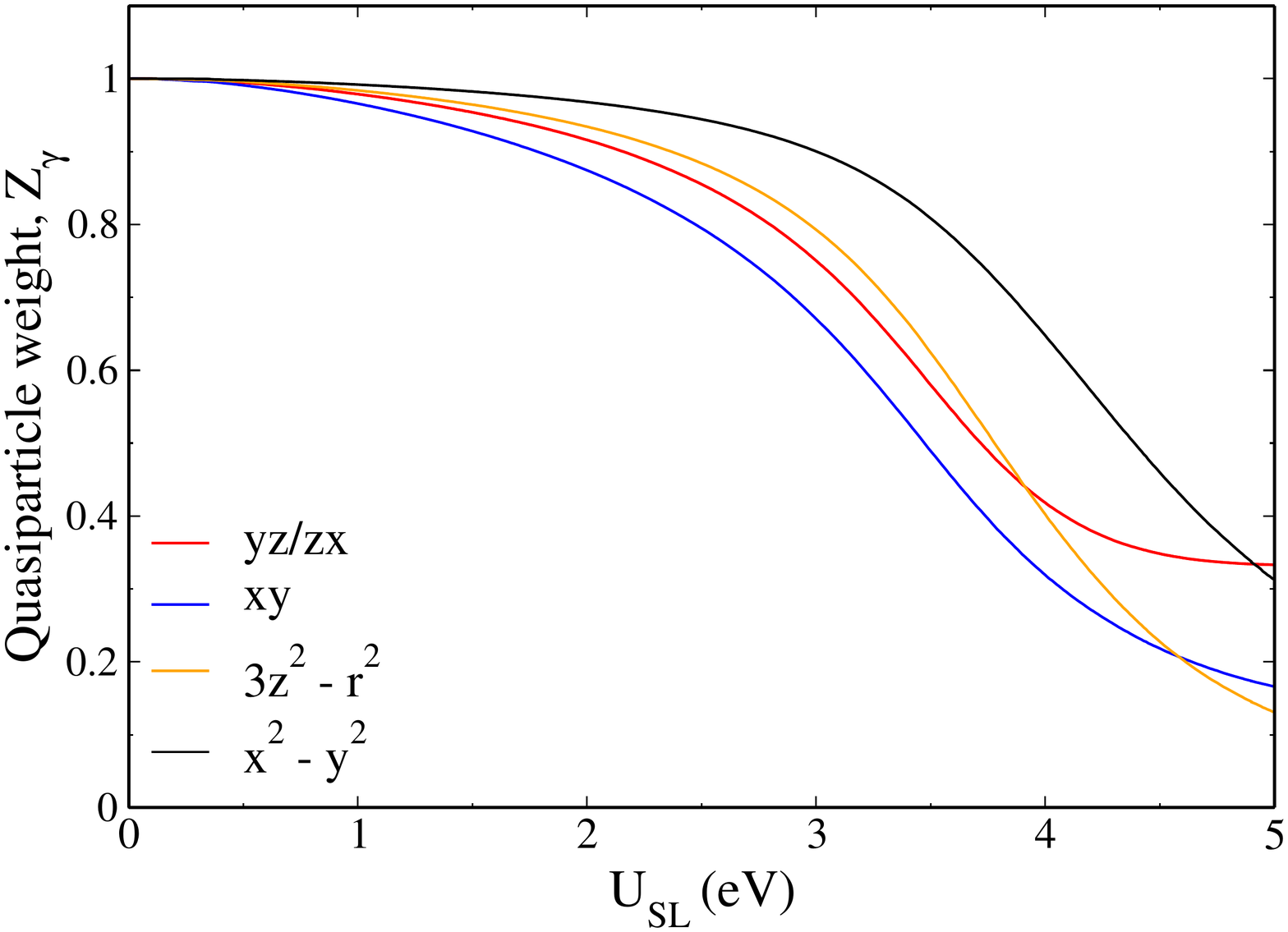}
\caption{Quasiparticle weights $Z_{\gamma}$ as a function of $U_{\rm SL}$ for the tight-binding corresponding to LaFeAsO at $n=4$. }
\label{fig:Fig.S3}
\end{figure}

Fig.~\ref{fig:Fig.S3} shows the quasiparticle weights $Z_{\gamma}$ as a function of $U_{\rm SL}$ for the tight-binding corresponding to LaFeAsO at $n=4$. At this filling the orbitals $3z^2-r^2$ and $x^2-y^2$ are half-filled. At low values of $U_{\rm SL}$, these orbitals are not the most correlated ones 
but this changes upon increasing the interactions, with $3z^2-r^2$ becoming the most correlated orbital above $U_{\rm SL} \sim 4.5$ eV.

{\bf Multi-orbital RPA.}
To search for the leading superconducting instabilities we assume that the pairing interaction is due to the exchange of spin or orbital fluctuations. 
We focus on singlet solutions and use Random Phase Approximation (RPA) calculations to determine the momentum dependence of the gap function corresponding to the leading eigenvalue of the pairing interaction vertex, expected to give the highest critical temperature. The technique was developed in~\cite{takimoto_prb2004,kubo_prb2007} and later explained in detail within the context of iron superconductors~\cite{graser_njp2008} where it has been used extensively.

\bibliography{papern4}

\end{document}